\documentclass[aps,prb,twocolumn,floatfix,showpacs]{revtex4-1}
\usepackage{ifpdf}
\pdfoutput=1
\usepackage{graphicx}
\usepackage{subfigure}
\usepackage{amsmath,amssymb}% Include figure files
\usepackage{color}

\makeatletter

\usepackage{verbatim}

\makeatletter

\def\be{\begin{equation}}
\def\ee{\end{equation}}
\def\ba{\begin{eqnarray}}
\def\ea{\end{eqnarray}}

\makeatother

\begin{document}

\title{Effects of Long-Range Interactions on Magnetic Excitations and Phase Transition on a Magnetically Frustrated Square Lattice}

\author{Yuting Tan}
\author{Dao-Xin Yao}
\email{yaodaox@mail.sysu.edu.cn}

\affiliation{State Key Laboratory of Optoelectronic Materials and Technologies, School of Physics and Engineering, Sun Yat-sen University, Guangzhou 510275, China}
\date{\today}

\begin{abstract}
We investigate the effects of long-range interaction on the magnetic excitations and the competition between magnetic phases on a frustrated square lattice.
Applying the spin wave theory and assisted with symmetry analysis, we obtain analytical expression for spin wave spectrum of competing Neel and $(\pi, 0)$ stripe states of systems containing any-order long-range interactions. In the specific case of long-range interactions with power-law decay, we found surprisingly that staggered long-range interaction suppresses quantum fluctuation and enlarges the ordered moment, especially in the Neel state, and thus extends its phase boundary to the stripe state. Our findings only illustrate the rich possibilities of the roles of long-range interactions, and advocate future investigations in other magnetic systems with different structures of interactions.

%There is an adjustable phase transition from the $(\pi,\pi) $ antiferromagnet to the $(\pi, 0)$ antiferromagnet.
\end{abstract}
\maketitle
%semiclassical
%real materials
%Provide the reference for the QMC
%future work: Monte carlo

\section{Introduction}

Contrary to the well studied magnetic systems with short-range coupling, the physical effects of long-range magnetic interactions remain an important current research topic.
Practically, long-range interactions are common in magnetic systems, for example with metallic carriers that mediate the interactions~\cite{ohno,apl02}, in the form of the well-known RKKY interaction~\cite{rkky} or the double exchange interaction~\cite{yaoprl2011}.
For example, the RKKY interaction has also been found on the surface of three dimensional topological insulators~\cite{chenprl}, where the magnetic impurities are mediated by the helical Dirac electrons.~\cite{yaoprl2011}.
Another current heavily debated case is the magnetic properties of iron-based superconductors, which hosts clear signal of local moment and itinerant magnetic carriers~\cite{philip,zhang09}.
Obviously, in such a metallic system, if one were to integrate out the itinerant degree of freedom to obtain a spin only system, the interactions would be long-range as well.
From these large classes of materials of current interest, it is obvious that a better understanding of the effects of long-range magnetic interaction is of great scientific interest and practical importance.
This is particularly so when the systems contains frustrated short-range interactions and competing phases.

Previously, the effects of long-range interactions have been studied in one-dimensional systems, and found to induce various interesting phenomena, including long-range order, quasi-long range order, valence bond solid phase,etc.~\cite{aoki95,yusuf04,affleck05,sandvik10}
On the other hand, the study of effects of long-range interaction remains largely unaddressed, mainly because of technical problems.

In this paper, we investigate the effects of long-range magnetic interaction in a square lattice containing frustrated first- and second-neighbor interactions.
Assisted with symmetry analysis, we derive a general analytical expression for the spin wave spectrum in the competing Neel and $(\pi,0)$ stripe states.
We then study their phase competition under the influence of staggered long-range interaction with power-law decay.
We found that the long-range interaction widens the spin wave spectrum, enlarges the stiffness of the spin wave, and thus suppresses the quantum fluctuation of the spin, giving rise to a larger ordered moment, particularly in the Neel state.
This in turn strengthens the Neel state and extends its phase boundary to the stripe state.
Our results reveal the interesting physical effects of long-range interaction in this specific frustrated system, and offer a starting point for future systematic investigations on the important role of long-range interactions in frustrated magnetic systems in general.

This paper is organized as follows. In Section II we begin with a
brief description of model and method. In Section III we present the
general analytical results of spin waves for both the ($\pi, \pi$)
Neel phase and ($\pi, 0$) stripe phase with any-order long-range interactions.
In Section IV we study an adjustable long-range interaction with the
form of $1/|r|^{\alpha}$. The spin wave dispersions, dynamic structure
factor, constant energy slices, reduced magnetic moment and phase diagrams
are obtained. Finally, in Section V we summarize the main results.

\section{Model and Method}

%%%%%%%%%%%% FIGURE %%%%%%%%%%%%%%%%%%%%%%%%%%%%%%%%%%%
\begin{figure}[h]
{\centering
  \subfigure[]
  {\resizebox*{!}{0.7\columnwidth}{
  \includegraphics{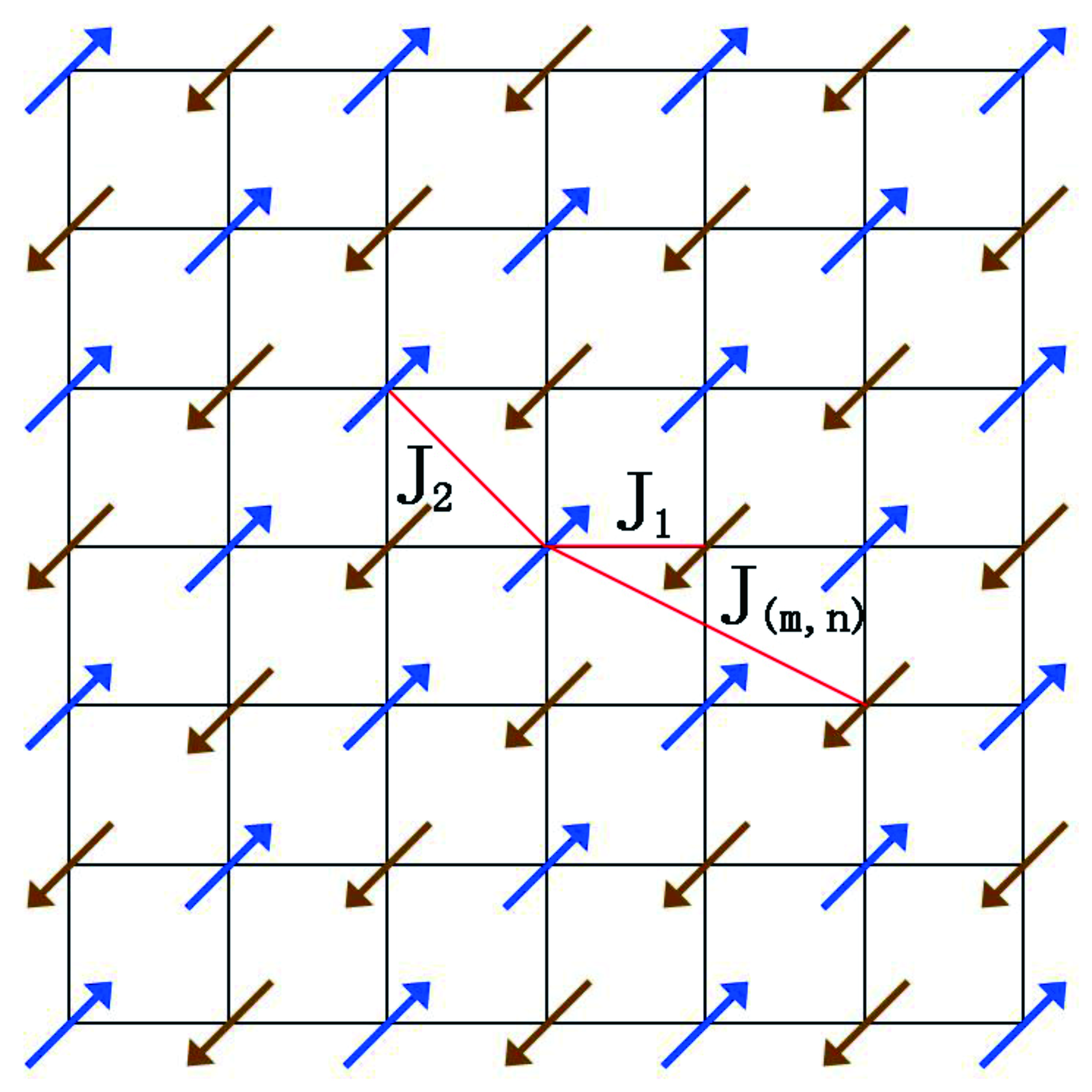}\label{neel phase lattice}}}
  \subfigure[]
  {\resizebox*{!}{0.7\columnwidth}{
   \includegraphics{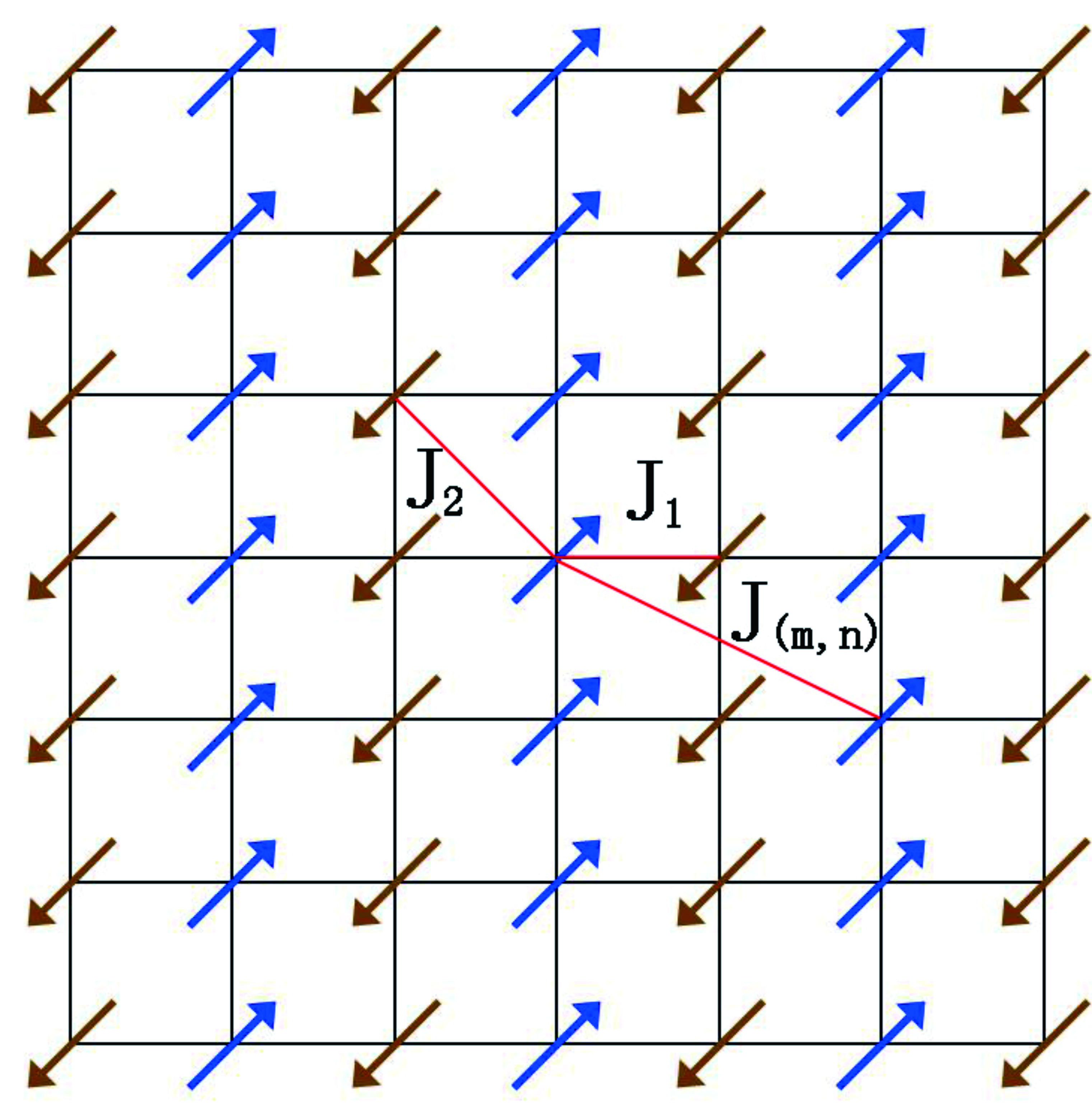}\label{stripe phase lattice}}}
  \par}
\caption{Magnetic ground states with long-range interactions: (a)
Neel phase with wave vector ($\pi, \pi$) and (b) Stripe phase with
wave vector ($\pi, 0$). $J_{(m, n)}$ represents the interactions
between two spins with relative coordinates $(m, n)$.}
\label{lattice}
\end{figure}
%%%%%%%%%%%%%%%%%%%%%%%%%%%%%%%%%%%%%%%%%%%%%%%%%%%%%%%

%Hamitonian

We study an extended version of the usual nearest-neighbor Heisenberg model on the two-dimensional square lattice.
%by using a Heisenberg model with long-range interactions. %By tuning the interactions, the system can be in the Neel ($\pi, \pi$) phase or the stripe ($\pi, 0$) phase.
The Hamiltonian is written as
\begin{equation}
H = \sum_{<\mathbf{r},\mathbf{r'}>} J_{\mathbf{r},\mathbf{r'}} \mathbf{S}_{\mathbf{r}} \cdot \mathbf{S}_{\mathbf{r'}},
\label{model}
\end{equation}
where \(<\mathbf{r},\mathbf{r'}>\) are spin sites, and \(J_{\mathbf{r},\mathbf{r'}}\) is the exchange coupling. To describe the long-range interactions conveniently, we use
$J_{\mathbf{r},\mathbf{r'}}=J_{(m, n)}$ to represent the interaction between two spins with
the relative coordinates $(m, n)$ shown in Fig.~\ref{lattice}. For example, if we choose one site as $(0, 0)$, then its neighboring sites can be coordinated by $(m, n)$, where $m$ is the relative $x$-coordinate and $n$ is the relative $y$-coordinate to the $(0, 0)$-point. From this definition, we can represent $J_1=J_{(1, 0)}$, $J_2=J_{(1, 1)}$, $J_3=J_{(2, 0)}$, and so on. Also we have $J_{(|m|, |n|)}=J_{(|n|, |m|)}$ by the symmetry. Therefore, we can rewrite the Hamitonian as
\begin{eqnarray}
H&= &\sum_{<\mathbf{r},\mathbf{r'}>}J_{(m,n)} \mathbf{S}_{\mathbf{r}}\cdot \mathbf{S}_{\mathbf{r'}},
\end{eqnarray}
where $J_{(m,n)}$ depends on the relative coordinates between $\mathbf{S}_{\mathbf{r}}$ and $\mathbf{S}_{\mathbf{r'}}$.
%Specially,$J_2$ are competing next-nearest neighbor interactions, and$ J_2^{'} $are long-range interactions.

By tuning the interactions, the system can be in the $(\pi,\pi)$
antiferromagnetic phase (see Fig.~\ref{neel phase lattice}) and
$(\pi,0)$ antiferromagnetic phase (see Fig.~\ref{stripe phase
lattice}). The $(\pi,\pi)$ antiferromagnetic phase is the ground state of undoped cuprates, and $(\pi,0)$ antiferromagnetic phase is closely related to iron pnictides.~\cite{dai08,zhao09,ma09}
There are two spins in each unit cell for these two
antiferromagnetic phases.

%spin wave theory

We use Holstein-Primakoff bosons to quantize about the
antiferromagnetic ground states.
\begin{equation}
  H=E_{Cl}+S\sum_{\mathbf{k}} [A_{\mathbf{k}}
  a_{\mathbf{k}}^+a_{\mathbf{k}}+\frac{1}{2}(B_{\mathbf{k}}
  a_{\mathbf{k}}^+a_{-\mathbf{k}}^++B_{\mathbf{-k}}^*
  a_{\mathbf{k}}a_{-\mathbf{k}}]
\end{equation}
where $E_{Cl}$ is the classical ground state energy which depends on the spin configuration and interactions.

The Hamiltonian can be diagonalized using the Bogoliubov
transformation~\cite{erica04}
\begin{equation}
  b_{\mathbf{k}}=\cosh{\theta_{\mathbf{k}}a_{\mathbf{k}}}-\sinh{\theta_{\mathbf{k}}} a_{-\mathbf{k}}^+.
\end{equation}
The diagonalized Hamiltonian is
\begin{equation}
H=\sum_{\mathbf{k}}\omega(\mathbf{k})b_{\mathbf{k}}^+b_{\mathbf{k}}+E_{Cl}+E_{0}
\end{equation}
where $\omega(\mathbf{k})$ is the spin wave dispersion
\begin{equation}
\omega(\mathbf{k})= S\sqrt{ A_{\mathbf{k}}^2- B_{\mathbf{k}}^2},
\end{equation}
and $E_0$ is the quantum zero-point energy correction
\begin{equation}
E_0=\frac{S}{2}\sum_{\mathbf{k}}(-A_{\mathbf{k}}+\omega(\mathbf{k})).
\end{equation}

%structure factor

The dynamic structure factor $S(\mathbf{k},\omega)$ is an important
quantity which is proportional to the neutron scattering cross section.
In the linear spin-wave approximation, only the transverse parts
contribute to the dynamic structure factor. We have
\begin{eqnarray}
S^{xx}(\mathbf{k},\omega)&=&S^{yy}(\mathbf{k},\omega)\nonumber\\
&=&g^2\mu_B^2S\frac{A_k-B_k}{2\omega(\mathbf{k})}[n(\omega)+1]\delta(\omega-\omega(\mathbf{k})),\nonumber\\
\label{strfactor}
\end{eqnarray}
where $g$ is the $g$-factor, and $n(\omega)$ is the Bose occupation factor.~\cite{yaofront09,boothroyd08}

%The integrated structure factor $S(\omega)$ can also be used to
%analyze the neutron scattering data
%\begin{equation}
%S(\omega)^{\alpha\alpha}=\int\int\int_{BZ}dk_xdk_ydk_zS^{\alpha\alpha}(\mathbf{k},\omega)\delta(\omega-\omega(\mathbf{k})),
%\end{equation}
%where $\alpha=x,y$ and BZ represents the full magnetic Brillouin zone.

\section{Analytical Results}

%tyt
%改动 shown in Fig.~\ref{neel phase lattice}
Through the linear spin wave theory as mentioned in Section II, we
obtain the analytical results for both the Neel phase and stripe
phase with long-range interactions.  The spin wave dispersion is given by
%提一下铜氧超导体 材料

\begin{equation}
\omega(\mathbf{k})= S\sqrt{ A_{\mathbf{k}}^2- B_{\mathbf{k}}^2},
\label{disperationf}
\end{equation}
where

\begin{eqnarray}
   A_{\mathbf{k}} &=& \sum_{(m,n)}A_{(m,n)},
   \label{disperation1}\\
   B_{\mathbf{k}} &=& \sum_{(m,n)}B_{(m,n)}.
   \label{disperation2}
\end{eqnarray}

Here $A_{(m,n)}$ and $B_{(m,n)}$ depend on the symmetry of ground state and will be given
separately.

\subsection{$(\pi,\pi)$ Neel phase}

According to the geometric structure of $(\pi,\pi)$ antiferromagnet with any order long-range interactions,
we find that

\begin{eqnarray}
   A_{(m,n)} &=& 4{(-1)}^{m+n+1}\alpha(m,n)J_{(m,n)}\nonumber\\
             &+& [{(-1)}^{m+n}+1]\alpha(m,n)J_{(m,n)}\nonumber\\
             &&[\cos{(mk_x)}\cos{(nk_y)}+\cos{(nk_x)}\cos{(mk_y)}],\nonumber\label{neel1}\\ \\
   B_{(m,n)} &=& [{(-1)}^{m+n+1}+1]\alpha(m,n)J_{(m,n)}\nonumber\\
             &&[\cos{(mk_x)}\cos{(nk_y)}+\cos{(nk_x)}\cos{(mk_y)}].\nonumber\label{neel2}\\
\end{eqnarray}

Here we have defined
\begin{equation}
\alpha(m,n)= \left\{
\begin{array}{*{20}{c}}
&1& &{m=0}~or~{n=0}~or~{m=n} \\
&2& &{otherwise}
\end{array}
\right.
\end{equation}

The coefficients in Eqs.~(\ref{neel1}) and (\ref{neel2}) are determined by the direction of spins. In the Neel phase, we find that $(m+n)$ is always even if the spin at $(m, n)$ has the same direction with the one at $(0,0)$, otherwise it is odd. $\alpha(m,n)$ is $1$ if the spins are located at the high symmetry points, i.e. ${m=0}~or~{n=0}~or~{m=n}$; otherwise it is $2$.

 %$(-1)^{m+n+1}$ is negative when the summation of its coordinate $m+n$ is even in the Neel phase, the spin is up a
 %On the contrary, $(-1)^{m+n+1}$ is positive and the spin is down for the case $m+n$ is an odd number.

%When the spin is up, the corresponding term only occur in $A_{(m,n)}$ and $(-1)^{m+n}+1=2$, but it only occur in $B_{(m,n)}$ when the spin is down and $(-1)^{m+n+1}+1=2$.

In the case of $|m|=1$ and $|n|=0$, we recover the classical result for the $J_1$-only model on the square lattice ~\cite{kruger}
\begin{eqnarray}
   A_{\mathbf{k}} &=& 4 J_1\\
   B_{\mathbf{k}} &=& 2 J_1[\cos{(k_x)}+\cos{(k_y)}]
   \label{disperationmn2}
\end{eqnarray}

In the case of $|m|\leq1$ and $|n|\leq1$, which has no long-range
interaction, we have
\begin{eqnarray}
   A_{\mathbf{k}} &=& 4(J_1-J_2)+4J_2\cos{(k_x)}\cos{(k_y)}\\
   B_{\mathbf{k}} &=& 2J_1[\cos{(k_x)}+\cos{(k_y)}]
   \label{disperationmn1}
\end{eqnarray}

From this expression, we can see that the spin wave dispersion is invariant if we switch $k_x$ and $k_y$ even the long-range interactions are included. This reflects the symmetry of $(\pi, \pi)$ phase.

\subsection{$(\pi,0)$ Stripe phase}

For the $(\pi,0)$ antiferromagnet with long-range interactions, the spin wave dispersion is also
given by Eqs.~(\ref{disperationf}),(\ref{disperation1})and(\ref{disperation2}).
 We have found

\begin{eqnarray}
A_{(m,n)} &=& 2{(-1)}^{m+1}\vert{(-1)}^m+{(-1)}^n\vert\alpha(m,n)J_{(m,n)}\nonumber\\
          &+& \alpha(m,n)J_{(m,n)}\{[{(-1)}^m+1]\cos{(mk_x)}\cos{(nk_y)}\nonumber\\
          &+&[{(-1)}^n+1]\cos{(nk_x)}\cos{(mk_y)}\},\label{stripe1}\\
B_{(m,n)} &=& \alpha(m,n)J_{(m,n)}\{[{(-1)}^{m+1}+1]\cos{(mk_x)}\cos{(nk_y)}\nonumber\\
          &+&[{(-1)}^{n+1}+1]\cos{(nk_x)}\cos{(mk_y)}\}.\label{stripe2}
\end{eqnarray}

%The coefficient in Eqs.~(\ref{stripe1})and (\ref{stripe2})obey the same rule as neel phase. It can be known
 %that the different shape of spin wave dispersion comes from the symmetry of different ground state.
In the stripe phase, $m$ is even if the spin at $(m, n)$ has the same direction with the one at $(0, 0)$, otherwise it is odd. Its spin wave band is quite different from the Neel phase.
 %that the different shape of spin wave dispersion comes from the symmetry of different ground state.
%%tyt

When considering the case with $|m|\leq1$ and $|n|\leq1$, we can get
\begin{eqnarray}
   A_{\mathbf{k}} &=&2J_1\cos{(k_y)}+4J_2\\
   B_{\mathbf{k}} &=& 2J_1\cos{(k_x)}+4J_2\cos{(k_x)}\cos{(k_y)},
   \label{disperationmn2}
\end{eqnarray}
which is exactly the same result of $J_1$-$J_2$ model.~\cite{yao08J1J2,si08,fang08}

This spin wave dispersion is asymmetric if we switch $k_x$ and
$k_y$. This reflects the symmetry of $(\pi, 0)$ phase.

To get the dynamic structure factor, we just need to substitute the
$A_K$ and $B_k$ obtained above into Eq.~(\ref{strfactor}).

%We can obtain the spin wave dispersions by simply substituting the parameters into the above general solutions.
With the above analytic solutions, we can easily explore the spin wave dispersions, spin wave velocities, dynamic structure factor, reduced magnetic moment
and phase transitions for the $(\pi,\pi)$ Neel antiferromagnet and $(\pi, 0)$ stripe antiferromagnet with all kinds of long-range interactions.
In the following Section, we will use an adjustable power-law long-range interaction to study the magnetic excitations and phase transition.

\section{Power-law long-range interactions}
%%%%%%%%%%%% FIGURE %%%%%%%%%%%%%%%%%%%%%%%%%%%%%%%%%%%
\begin{figure}[t]
{\centering
  \subfigure[]
  {\resizebox*{!}{0.6\columnwidth}{
  \includegraphics{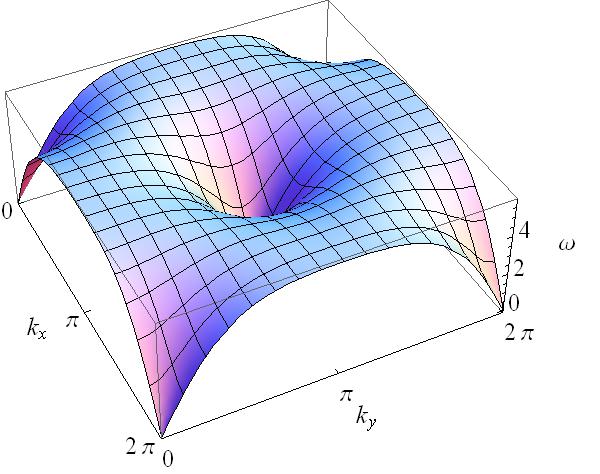}\label{spinwave401}}}
  \subfigure[]
  {\resizebox*{!}{0.6\columnwidth}{\includegraphics{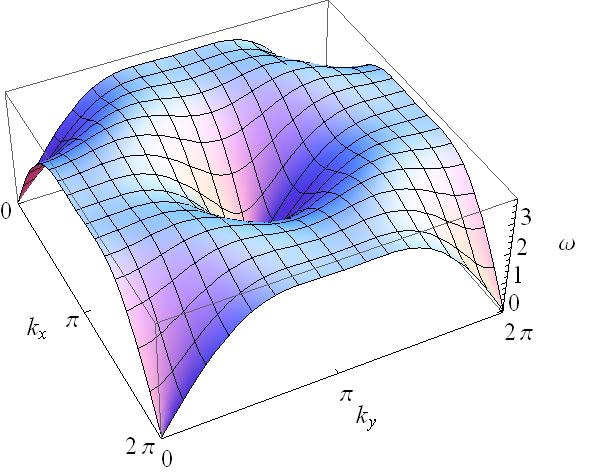}\label{spinwave-1001}}}
  \par}
\caption{(Color online) Spin-wave dispersion band for the
$(\pi,\pi)$ Neel phase with $J_2=0.1$, $J_1=1$: (a) $\alpha=4$ (long-range interactions) and (b) $\alpha=10$ (short range interactions).}
\label{Neel phase spinwave}
\end{figure}
%%%%%%%%%%%%%%%%%%%%%%%%%%%%%%%%%%%%%%%%%%%%%%%%%%%%%%%

%%%%%%%%%%%% FIGURE %%%%%%%%%%%%%%%%%%%%%%%%%%%%%%%%%%%
\begin{figure}[t]
{\centering
  \subfigure[]
  {\resizebox*{!}{0.6\columnwidth}{
  \includegraphics{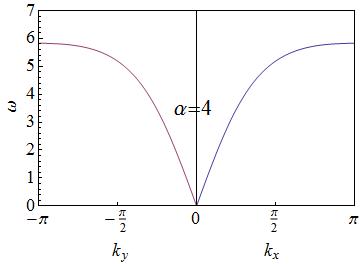}\label{velocity401}}}
  \subfigure[]
  {\resizebox*{!}{0.6\columnwidth}{\includegraphics{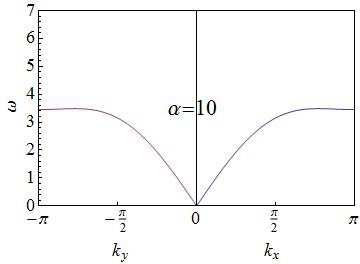}\label{velocity1001}}}
  \par}
\caption{(Color online) Spin-wave dispersions for the $(\pi,\pi)$
Neel phase with $J_2=0.1$, $J_1=1$: (a) $\alpha=4$ (long-range interactions and
(b) $\alpha=10$ (short range interactions). The spin wave velocities $v_x$ and $v_y$ increase about $80\%$ from $\alpha=10$ to $4$.} \label{Neel phase
velocity}
\end{figure}
%%%%%%%%%%%%%%%%%%%%%%%%%%%%%%%%%%%%%%%%%%%%%%%%%%%%%%%
In this section, we study the long-range interactions which decay as
a power law $1/|r|^{\alpha}$:
\begin{eqnarray}
J_{(m,n)}& =&{\left( { - 1} \right)^{\left| m \right| + \left| n \right| + 1}}\frac{{\lambda {J_1}}}{{{{|r|}^\alpha }}} \nonumber\\
&=& {\left( { - 1} \right)^{\left| m \right| + \left| n \right| + 1}}\frac{{\lambda {J_1}}}{{{{(\sqrt {{m^2} + {n^2}} )}^\alpha }}},
\label{def}
\end{eqnarray}
where $\lambda$ denotes the relative strength compared to $J_1$ and
$\alpha$ is the power-law exponent that controls the decay of
interactions. The factor $(-1)^{|m|+|n|+1}$ ensures that the
interactions are not frustrated.~\cite{aoki95,yusuf04}
Experimentally, this kind of power-law long-range interactions can
be realized by the RKKY interactions where local magnetic ions are
mediated by the itinerant electrons.~\cite{rkky} Theoretically, this
type of long-range interactions can be simulated by quantum monte
carlo methods without the notorious sign problem.~\cite{sandvik10} In this Section,
we use a circle of spins (61) which satisfy $\sqrt{m^2+n^2}\leq 3\sqrt{2}$ as an illustration. We also calculated the cases for $\sqrt{m^2+n^2}>3 \sqrt{2}$, which shows no difference when $\alpha \geq 2$.

Shown in Fig.~\ref{lattice}, we have $r={(\sqrt {{m^2} + {n^2}}
)}$. The corresponding interactions can be obtained from Eq.~(\ref{def}). Here, we use
\begin{equation}
J_2^{'} = J_2 -\frac{\lambda J_1}{\sqrt2^\alpha},
\end{equation}
where a NNN interaction $J_2$ is added together with the power-law
interaction to introduce the competition. The new $J_2^{'}$ includes
both $J_2$ and power-law term, which can be substituted into
previous formulas. We will see that $J_2$ is an important parameter
to control the phase transition. In the following, we will use
$\lambda=1$ for calculations.

\subsection{$(\pi, \pi)$ Neel phase}
%%%%%%%%%%%% FIGURE %%%%%%%%%%%%%%%%%%%%%%%%%%%%%%%%%%%
\begin{figure}[t]
{\centering
  \subfigure[]
  {\resizebox*{!}{0.6\columnwidth}{
  \includegraphics{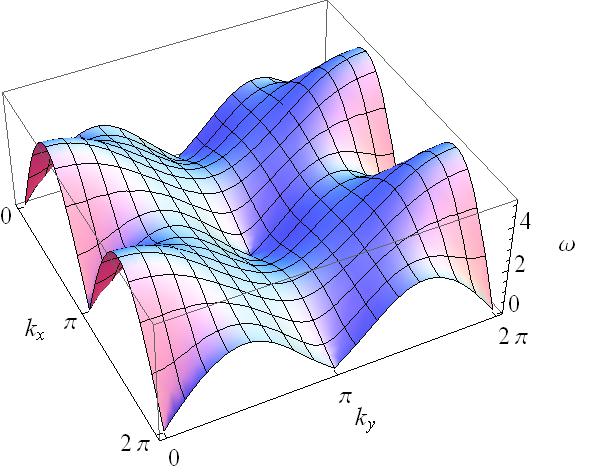}\label{spinwave41}}}
  \subfigure[]
  {\resizebox*{!}{0.6\columnwidth}{
  \includegraphics{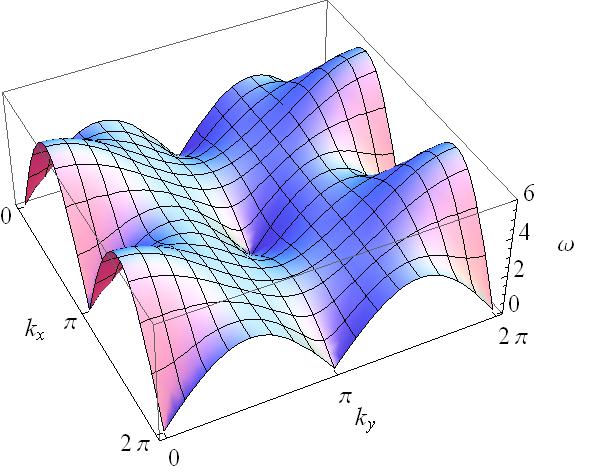}\label{spinwave101}}}
  \par}
\caption{(Color online) Spin-wave dispersion band for the $(\pi,0)$
stripe phase with $J_2=1$, $J_1=1$: (a) $\alpha=4$ (long-range
interactions) and (b) $\alpha=10$ (Short range interactions).}
\label{Stripe phase spinwave}
\end{figure}
%%%%%%%%%%%%%%%%%%%%%%%%%%%%%%%%%%%%%%%%%%%%%%%%%%%%%%%
%%%%%%%%%%%% FIGURE %%%%%%%%%%%%%%%%%%%%%%%%%%%%%%%%%%%
\begin{figure}[t]
{\centering
  \subfigure[]
  {\resizebox*{!}{0.6\columnwidth}{
  \includegraphics{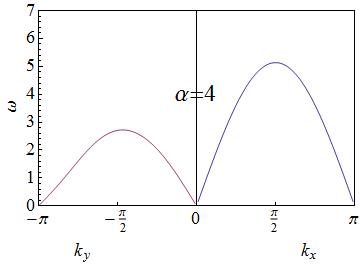}\label{velocity41}}}
  \subfigure[]
  {\resizebox*{!}{0.6\columnwidth}{\includegraphics{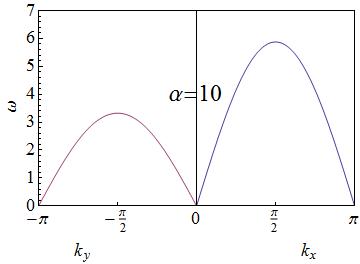}\label{velocity101}}}
  \par}
\caption{(Color online) Spin-wave dispersions for the $(\pi,0)$
stripe phase with $J_2=1$, $J_1=1$: (a) $\alpha=4$ (long-range interactions)
and (b) $\alpha=10$ (short range interactions). The spin wave velocity $v_x$ decreases about $13\%$ and $v_y$ drops about $20\%$ from $\alpha=10$ to $4$.} \label{Stripe phase
velocity}
\end{figure}
%%%%%%%%%%%%%%%%%%%%%%%%%%%%%%%%%%%%%%%%%%%%%%%%%%%%%%%
By substituting the above power-law long-range interactions into Eqs.~(\ref{disperation1}) and (\ref{disperation2}), we can get the spin wave dispersion

\begin{eqnarray}
A_{\mathbf{k}} &=& 4(\lambda J_1/\sqrt 2 ^\alpha -J_2) + 4(\lambda J_1/\sqrt 8 ^\alpha ) + 4(\lambda J_1/\sqrt {18} ^\alpha) \nonumber\\
&+& 4(\lambda J_1/2^\alpha ) +8(\lambda J_1/\sqrt {10} ^\alpha ) + 4J_1 + 8(\lambda J_1/\sqrt 5 ^\alpha )\nonumber\\
&+& 8(\lambda J_1/\sqrt {13} ^\alpha) + 4(\lambda J_1/3^\alpha )+4(\lambda J_1/4^\alpha) \nonumber\\
&+& 8(\lambda J_1/\sqrt{17}^\alpha)\nonumber\\
&-& 4(\lambda J_1/\sqrt 2 ^\alpha -J_2)\cos {(k_x)}\cos {(k_y)}\nonumber\\
&-& 4\lambda J_1/\sqrt 8 ^\alpha \cos {(2 k_x)}\cos {(2 k_y)}\nonumber\\
&-& 4\lambda J_1/\sqrt {18} ^\alpha \cos {(3 k_x)}\cos {(3 k_y)}\nonumber\\
&-& 2\lambda J_1/2^\alpha [\cos {(2 k_x)} + \cos {(2 k_y)}]\nonumber\\
&-&  4\lambda J_1/\sqrt {10} ^\alpha[\cos {(3k_x)}\cos {(k_y)}+ \cos {(k_x)}\cos{(3k_y)}]\nonumber \\
&-& 2\lambda J_1/4^\alpha[\cos{(4k_x)} + \cos{(4k_y)}]
\\
B_{\mathbf{k}}&=&  2J_1[\cos{(k_x)}+\cos{(k_y)}]+ 2\lambda J_1/3^\alpha [\cos{(3k_x)}+\cos{(3k_y)}]\nonumber\\
&+& 4\lambda J_1/\sqrt5^\alpha[\cos{(2k_x)}\cos{(k_y)}+\cos{(k_x)}\cos{(2k_y)}]\nonumber\\
&+& 4\lambda J_1/\sqrt{13}^\alpha[\cos{(3k_x)}\cos{(2k_y)}+\cos{(2k_x)}\cos{(3k_y)}]\nonumber\\
&+& 4\lambda J_1/\sqrt{17}^\alpha[\cos{(4k_x)}\cos{(k_y)} + \cos{(k_x)}\cos{(4k_y)]}
\end{eqnarray}

Fig.~\ref{Neel phase spinwave} shows the spin wave band with
different interactions for the $(\pi,\pi)$ Neel phase. The spin wave
band with long-range interactions is shown in
Fig.~\ref{spinwave401} for $J_2=0.1$ and $\alpha=4$. The other one with short range interactions is shown in
Fig.~\ref{spinwave-1001} for $J_2=0.1$ and $\alpha=10$. From the plots, we can see that the low energy spin wave bands are almost invariant for the two kinds of interactions. Both bands have $\omega\rightarrow 0$ at $(\pi,\pi)$ point which corresponds to the
magnetic wave vector. However, the difference shows up at high
energies: different band shapes and energy scales. This reflects the
geometry of interactions.

The associated spin wave velocities are
\begin{equation}
v_x = v_y = 2\sqrt{2}S\sqrt {(J_1 - 2J_2 + a_1\lambda J_1 )(J_1 +b_1\lambda J_1)},
\end{equation}
where\\
$\begin{array}{l}
{a_1} = {2^{1 - \frac{\alpha }{2}}} + {2^{2 - \alpha }} + {2^{3 - \frac{{3\alpha }}{2}}} + {3^{2 - \alpha }} + {2^{1 - \frac{\alpha }{2}}}\times {3^{2 - \alpha }} \\ ~~~+ 2 \times {5^{1 - \frac{\alpha }{2}}} + {2^{2 - \frac{\alpha }{2}}}\times {5^{1 - \frac{\alpha }{2}}} + 2\times{13^{1 - \frac{\alpha }{2}}}+2\times{17^{1 - \frac{\alpha }{2}}}\\
{a_2} = {3^{-\alpha }} + 2\times{5^{-\frac{\alpha }{2}}} + 2 \times{13^{-\frac{\alpha} {2}}}+ 2 \times{17^{-\frac{\alpha} {2}}}
\end{array}$

When $\alpha \rightarrow \infty$, we have %$a_1,b_1\rightarrow 0$, and
\begin{equation}
v_x = v_y = 2\sqrt{2}S\sqrt {J_1(J_1-2J_2)},
\end{equation}

From this equation, we can see that there is a phase transition at $J_2=0.5J_1$ for the $J_1$-$J_2$ model.

The spin wave velocities along $k_x$- and $k_y$-directions are the same because of the symmetry of $(\pi, \pi)$ phase. Fig.~\ref{Neel phase
velocity} shows the velocities (slope) along $k_x$- and $k_y$-directions at the point $(0,0)$ in the $k$ space.
It can be seen that the long-range interactions increase the spin wave velocities $v_x$ and $v_y$ dramatically in the
$(\pi,\pi)$ Neel phase, see Fig.~\ref{Neel phase velocity}.

\subsection{$(\pi,0)$ Stripe phase}

The stripe phase by definition is not only breaking the spin rotational symmetry but also breaks the crystal $C_4$ symmetry down to $C_2$.  %The $C4$ symmetry of the square 2D lattice is broken by the striped state down to $C2$, and thus this state has a nematic component.
%Taking this ground state, we
Substituting the same power law form of long-range interactions into Eqs.~(\ref{disperation1}) and
(\ref{disperation2}), we can get the spin wave dispersion for the $(\pi,0)$ stripe phase:

\begin{eqnarray}
A_{\mathbf{k}} &=& 4( J_2-\lambda J_1/\sqrt 2 ^\alpha) + 4\lambda J_1/\sqrt 8 ^\alpha  - 4\lambda J_1/\sqrt 18 ^\alpha\nonumber \\
&+& 4\lambda J_1/2^\alpha  - 8\lambda J_1/\sqrt {10} ^\alpha +4\lambda J_1/4^\alpha\nonumber\\
 &-& 4\lambda J_1/\sqrt 8 ^\alpha \cos{(2{k_x})}\cos{(2{k_y)}}\nonumber\\
 &-& 2\lambda J_1/2^\alpha[\cos{(2{k_x})} + \cos{(2{k_y})}]\nonumber\\
 &+& 2J_1\cos{(k_y)} + 4\lambda J_1/\sqrt{5} ^\alpha\cos{(k_y)}\cos{(2{k_x)}}\nonumber\\
 &+& 4\lambda J_1/\sqrt {13} ^\alpha\cos{(3k_y)}\cos{(2{k_x})}+ 2\lambda J_1/3^\alpha\cos{(3{k_y})} \nonumber\\
 &-& 2\lambda J_1/4^\alpha[\cos{(4k_x)} + \cos{(4k_y)}]\nonumber\\
 &+& 4\lambda J_1/\sqrt {17}^\alpha \cos{(k_y)}\cos{(4k_x)}  \\
B_{\mathbf{k}}&=&  4(J_2-\lambda J_1/\sqrt2^\alpha)\cos{(k_x)}\cos {(k_y)}\nonumber\\
 &-& 4\lambda J_1/\sqrt {18} ^\alpha \cos {(3k_x)}\cos {(3k_y)}\nonumber\\
 &-& 4\lambda J_1/\sqrt {10} ^\alpha [\cos{(3{k_x})}\cos {(k_y)} + \cos {(k_x)}\cos{(3{k_y)}}]\nonumber\\
 &+& 2{J_1}\cos{(k_x)} + 4\lambda J_1/\sqrt 5 ^\alpha \cos {(2{k_y)}}\cos {(k_x)}\nonumber\\
 &+& 4\lambda J_1/\sqrt {13} ^\alpha \cos{(2{k_y)}}\cos{(3{k_x})} + 2\lambda J_1/3^\alpha \cos{(3{k_x})} \nonumber \\
 &+& 4\lambda J_1/\sqrt {17}^\alpha\cos{(4k_y)}\cos{(k_x)}
\end{eqnarray}

Fig.~\ref{Stripe phase spinwave} shows the spin wave band for the
$(\pi,0)$ stripe phase. It is very different from the $(\pi, \pi)$
Neel phase because of the different symmetry.

Here we use $J_2=1$ in the $(\pi,0)$ stripe phase, which is much
larger than that in the $(\pi,\pi)$ Neel phase required by the
stability of ground state.  %Later, we will discuss the relationship between $J_2$ and the phase transition of two kind of ground state.
Comparing Fig.~\ref{spinwave41} with Fig.~\ref{spinwave101}, we can see that the long-range interactions here reduce the energy instead of increasing the energy in the $(\pi,\pi)$ Neel phase. It is because the long-range interactions here decrease the stability of $(\pi, 0)$ stripe phase. We will get to this point in the following part.

We can get the associated spin wave velocities
\begin{eqnarray}
v_x &=& 2S\{[2J_2 + J_1 (1 - a_3\lambda )]\nonumber\\
   &&(2J_2 + J_1 (1 - a_4\lambda))\}^{1/2}\nonumber\\
v_y &=& 2S\{[2J_2 + J_1 (a_5\lambda -1  )]\nonumber\\
& & (2J_2 + J_1 (1 - a_4\lambda))\}^{1/2}
\end{eqnarray}
where\\
$\begin{array}{l}
{a_3} =-2^{1 - \frac{\alpha}{2}} + 2^{2 -\alpha} + 2^{3-\frac{3\alpha}{2}} + 3^{2 -\alpha}- 2^{1 -\frac{\alpha}{2}}\times3^{2 -\alpha} \nonumber\\+ 4^{2 -\alpha} - 2^{2 -\frac{\alpha}{2}}\times 5^{1 -\frac{\alpha}{2}} -6 \times5^{-\frac{\alpha}{2}} + 10 \times13^{-\frac{\alpha}{2}} - 30 \times17^{-\frac{\alpha}{2}}\nonumber\\
{a_4} = 2^{1 -\frac{\alpha}{2}} - 3^{-\alpha} + 2^{1 -\frac{\alpha}{2}} \times3^{-\alpha} -
 2 \times5^{-\frac{\alpha}{2}} + 2^{2  -\frac{\alpha}{2}}\times 5^{-\frac{\alpha}{2}} \nonumber\\
 -2\times 13^{-\frac{\alpha}{2}} - 2 \times17^{-\frac{\alpha}{2}}\nonumber\\
{a_5} =-2^{1 - \frac{\alpha}{2}} + 2^{2 -\alpha} + 2^{3-\frac{3\alpha}{2}} + 3^{2 -\alpha}- 2^{1 -\frac{\alpha}{2}}\times3^{2 -\alpha} \nonumber\\+ 4^{2 -\alpha} - 2^{2 -\frac{\alpha}{2}}\times 5^{1 -\frac{\alpha}{2}} +6 \times5^{-\frac{\alpha}{2}} - 10 \times13^{-\frac{\alpha}{2}} + 30 \times17^{-\frac{\alpha}{2}}\nonumber\\
\end{array}$

When $\alpha \rightarrow \infty$, we can get %$a_2, b_2, c_2, a_3,b_3\rightarrow 0 $,
\begin{eqnarray}
v_x &=& 2S|2J_2+J_1|,\\
v_y &=& 2S\sqrt{4{J_2}^2-J_1^2}.
\end{eqnarray}
which are the results of $J_1$-$J_2$ model on the $(\pi, 0)$ stripe phase.~\cite{yao08J1J2} We find that the spin wave velocity $v_x$ decreases about $13\%$ and $v_y$ drops about $20\%$ from the short-range interaction case ($\alpha=10$) to the long-range interaction case $\alpha=4$).

The velocities along $k_x$- and $k_y$-directions are different and have more complicated
forms than that in the $(\pi,\pi)$ Neel phase. to the long-range interaction case  ($\alpha=10$).
First, the $(\pi,0)$ stripe phase is asymmetric along the $k_x$- and $k_y$-directions. The system is antiferromagnetic along $x$-direction and ferromagnetic along $y$-direction.
Second, we find that $v_x$ is always larger than
$v_y$, and the long-range interactions reduce the spin wave
velocities because they weaken the $(\pi, 0)$ stripe phase instead of
enhancing the $(\pi, \pi)$ Neel phase. Here only $J_2$ holds the $(\pi, 0)$ stripe phase.

%Yao

\subsection{Constant energy slices}
%%%%%%%%%%%% FIGURE %%%%%%%%%%%%%%%%%%%%%%%%%%%%%%%%%%%
\begin{figure}[t]
{\centering
  \subfigure[]
  {\resizebox*{!}{0.73\columnwidth}{
  \includegraphics{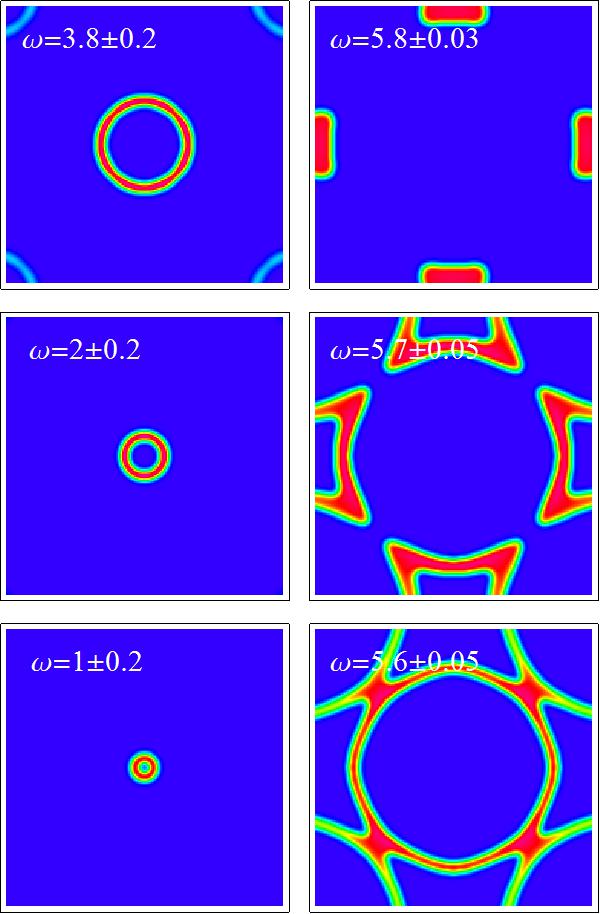}\label{phaseall401}}}
  \subfigure[]
  {\resizebox*{!}{0.73\columnwidth}{
  \includegraphics{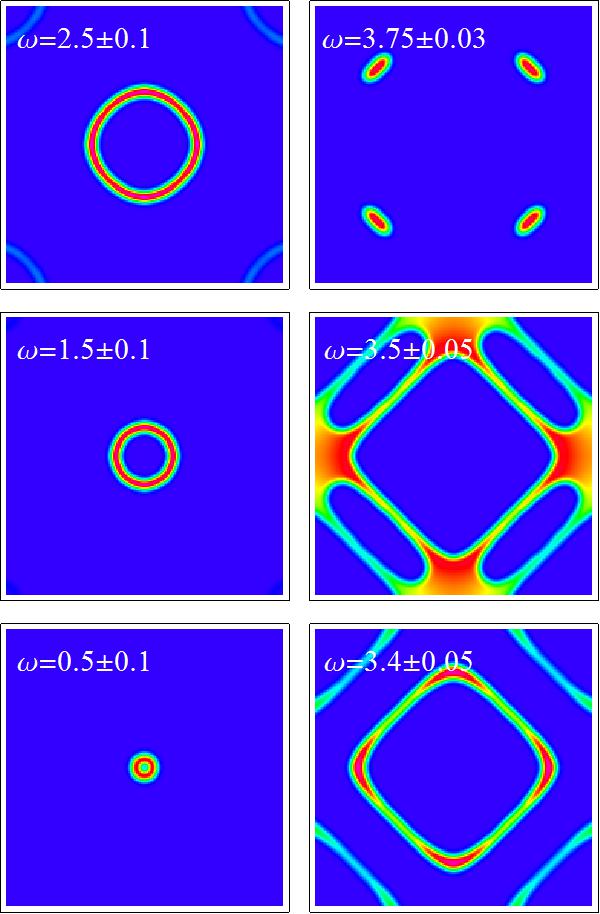}\label{phaseall1001}}}
  \par}
\caption{(Color online) Constant-energy slices (twinned) of the
dynamic structure factor $S(\mathbf{k},\omega)$ for $J_2=0.1$, $J_1=1$, $\lambda=1$: (a) $\alpha=4$ (long-range interactions) and
(b) $\alpha=10$ (short range interactions). The x-axis and y-axis
correspond to $k_x$ and $k_y$ respectively with the range $(0,
2\pi)$.} \label{Neel phase all}
\end{figure}
%%%%%%%%%%%%%%%%%%%%%%%%%%%%%%%%%%%%%%%%%%%%%%%%%%%%%%%
%%%%%%%%%%%% FIGURE %%%%%%%%%%%%%%%%%%%%%%%%%%%%%%%%%%%
\begin{figure}[t]
{\centering
  \subfigure[]
  {\resizebox*{!}{0.73\columnwidth}{
  \includegraphics{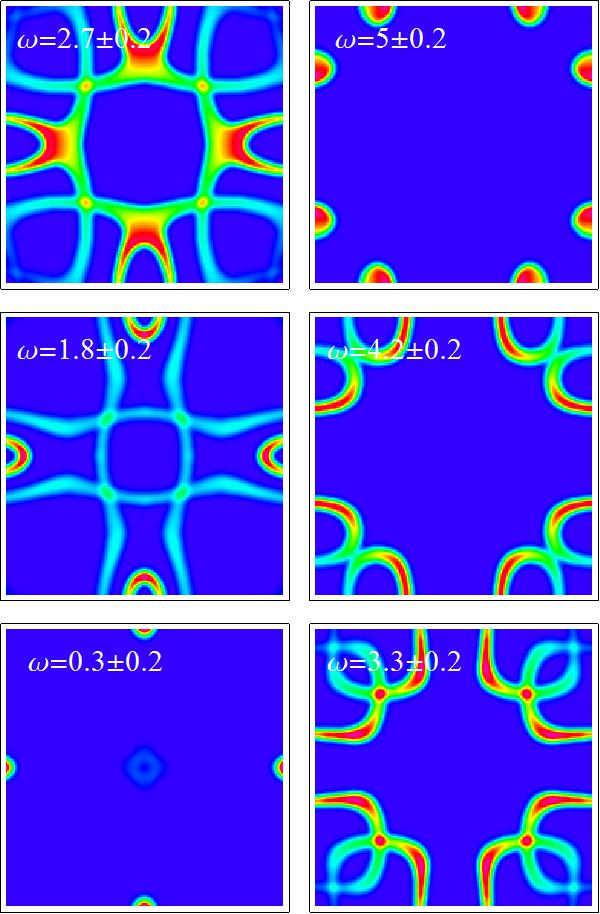}\label{phaseall41}}}
  \subfigure[]
  {\resizebox*{!}{0.73\columnwidth}{
  \includegraphics{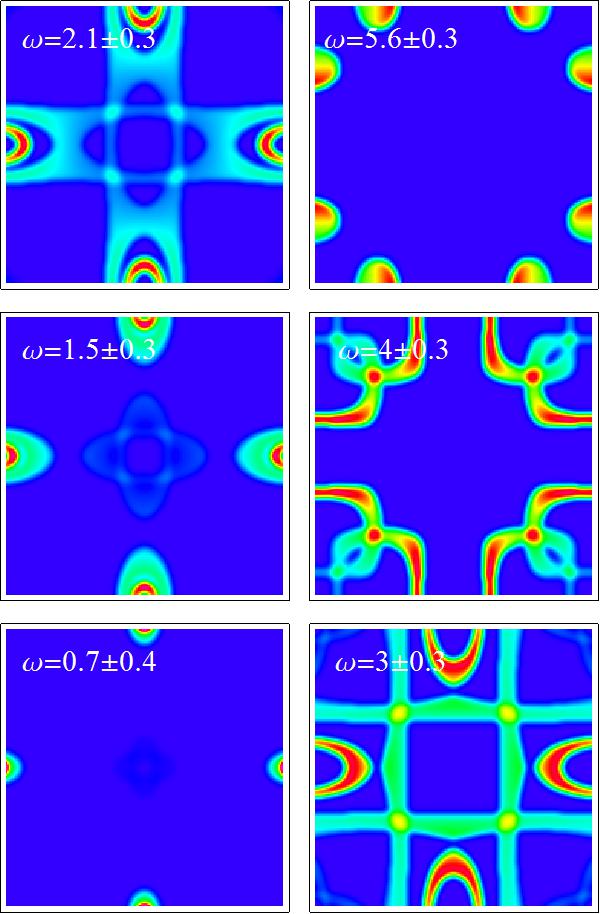}\label{phaseall101}}}
  \par}
\caption{(Color online) Constant-energy slices (twinned) of the
dynamic structure factor $S(\mathbf{k},\omega)$ for $J_2=1$, $J_1=1$, $\lambda=1$: (a) $\alpha=4$ (long-range interactions) and
(b) $\alpha=10$ (short range interactions). The x-axis and y-axis
correspond to $k_x$ and $k_y$ respectively with the range $(0,
2\pi)$.} \label{Stripe phase all}
\end{figure}
%%%%%%%%%%%%%%%%%%%%%%%%%%%%%%%%%%%%%%%%%%%%%%%%%%%%%%%

To compare with neutron scattering experiments, we calculate the constant energy slices. In Figs.~\ref{Neel phase all} and~\ref{Stripe phase all}, we show
the twinned neutron scattering intensity plots at constant energy
for the dynamic structure factor $S(\mathbf{k},\omega)$ in
$\mathbf{k}$-space, assuming a crystal with twinned $(\pi,\pi)$
antiferromagnetic domains (Fig.~\ref{Neel phase all}) or $(\pi,0)$
antiferromagnetic domains (Fig.~\ref{Stripe phase all}). In real
materials, spin order is generally twinned because of crystal
twining and local disorder pinning.~\cite{yao06,yao08a,boothroyd09} For this reason, we show the
twinned constant energy cutting plots which can be detected by
inelastic neutron scattering experiment.

For the $(\pi,\pi)$ Neel phase, there is a main peak
located at $(\pi, \pi)$ at low energy for both the long-range
interactions (Fig.~\ref{phaseall401}) and short range interactions
(Fig.~\ref{phaseall1001}), which corresponds to the magnetic wave
vector of Neel phase.  As energy increases, the peak increases
quickly to an outer ring. At higher energy, the ring forms bright spots and they touch each other.

However, there is a clear difference between the long-range
interactions and short range interactions at high
energies. For example, the central ring is almost a circle at high energy, while it is a square for the short range interactions. This is because the long-range interactions bring
more symmetry to the system. For the case with long-range interactions, the peaks are located at $(0,\pi)$, $(\pi,0)$,
$(2\pi,\pi)$ and $(\pi,2\pi)$. In the second case, the band tops are
located at $(\pi/2,\pi/2)$, $(\pi/2,3\pi/2)$, $(3\pi/2,\pi/2)$ and
$(3\pi/2,3\pi/2)$. In addition, we find that the long-range
interactions raises the top of energy while keeping other parameters
the same.

For the $(\pi,0)$ stripe phase, we show the constant energy
slices with $\alpha =4$ and $\alpha=10$ (Fig.~\ref{Stripe phase
all}). At low energy, there is one diffraction peak located at
$(\pi,0)$, which is the magnetic wave vector of the stripe phase. Unlike
the $(\pi, \pi)$ Neel phase, the low energy spin wave cones are generally
elliptical. At higher energies, the peaks are located at $(\pi/2,0)$, $(0,\pi/2)$ and the symmetry related points
(see Figs.~\ref{phaseall41} and ~\ref{phaseall101}). Contrary to the  We notice that the neutron scattering patterns are not sensitive to the long-range interactions. However, the long-range interactions can decrease the band top rather than increasing it because they try to destroy the $(\pi,0)$ phase. %the the different symmetry of these two phases.

\subsection{Reduced magnetic moment}
%%%%%%%%%%% FIGURE %%%%%%%%%%%%%%%%%%%%%%%%%%%%%%%%%%%
\begin{figure}[t]
\begin{center}
\resizebox*{0.9\columnwidth}{!}{\includegraphics{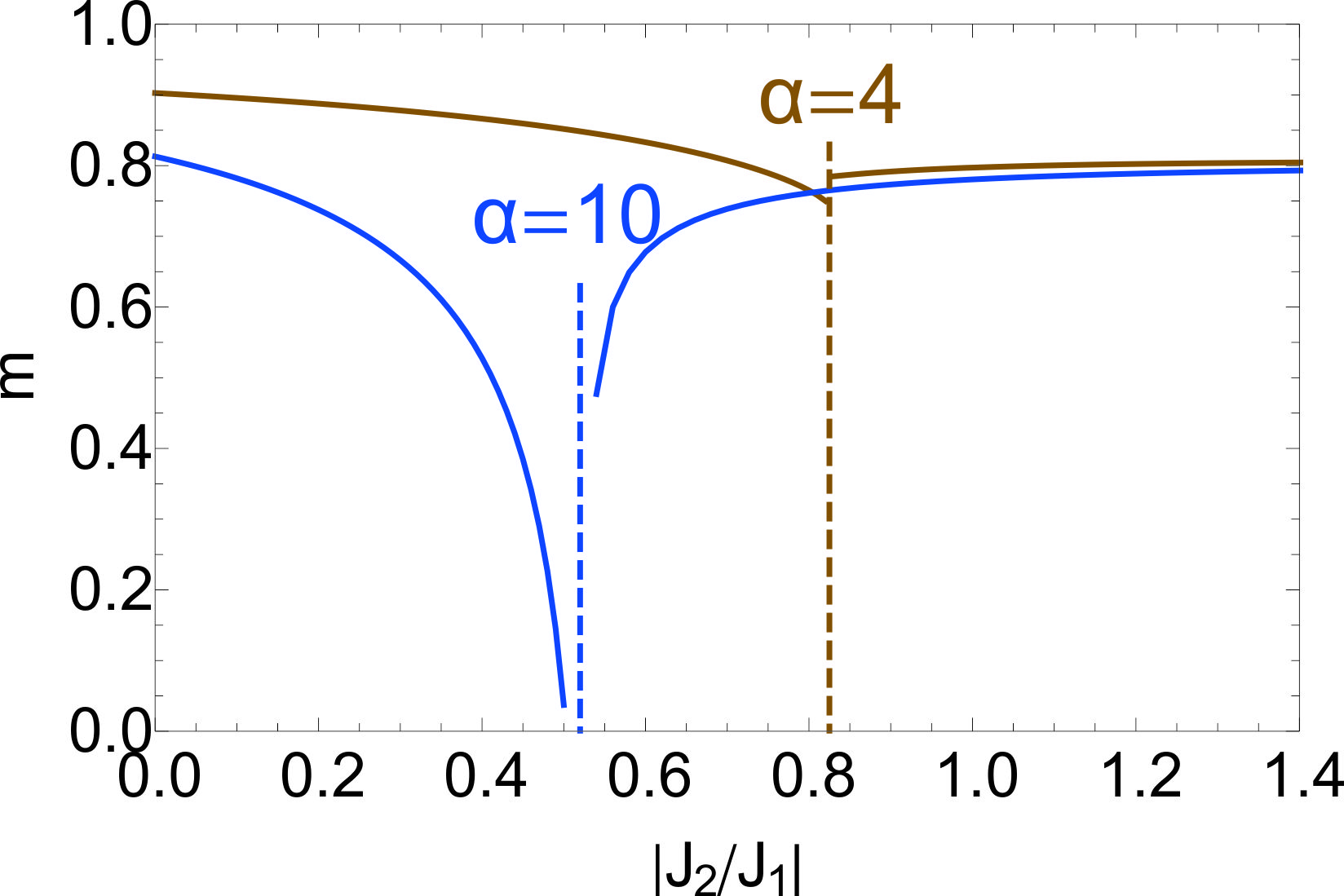}}
\end{center}
\caption{(Color online) $|J_2/J_1|$ dependence of $m$ for
$|J_1|=1$, $\lambda=1$ and $S=1$. Blue is for $\alpha=10$ (long-range
interactions) and brown is for $\alpha=4$ (short range interactions).}
\label{alltransition}
\end{figure}
%%%%%%%%%%%%%%%%%%%%%%%%%%%%%%%%%%%%%%%%%%%%%%%%%%%%%%
%%%%%%%%%%% FIGURE %%%%%%%%%%%%%%%%%%%%%%%%%%%%%%%%%%%
\begin{figure}[thb]
\begin{center}
\resizebox*{0.9\columnwidth}{!}{\includegraphics{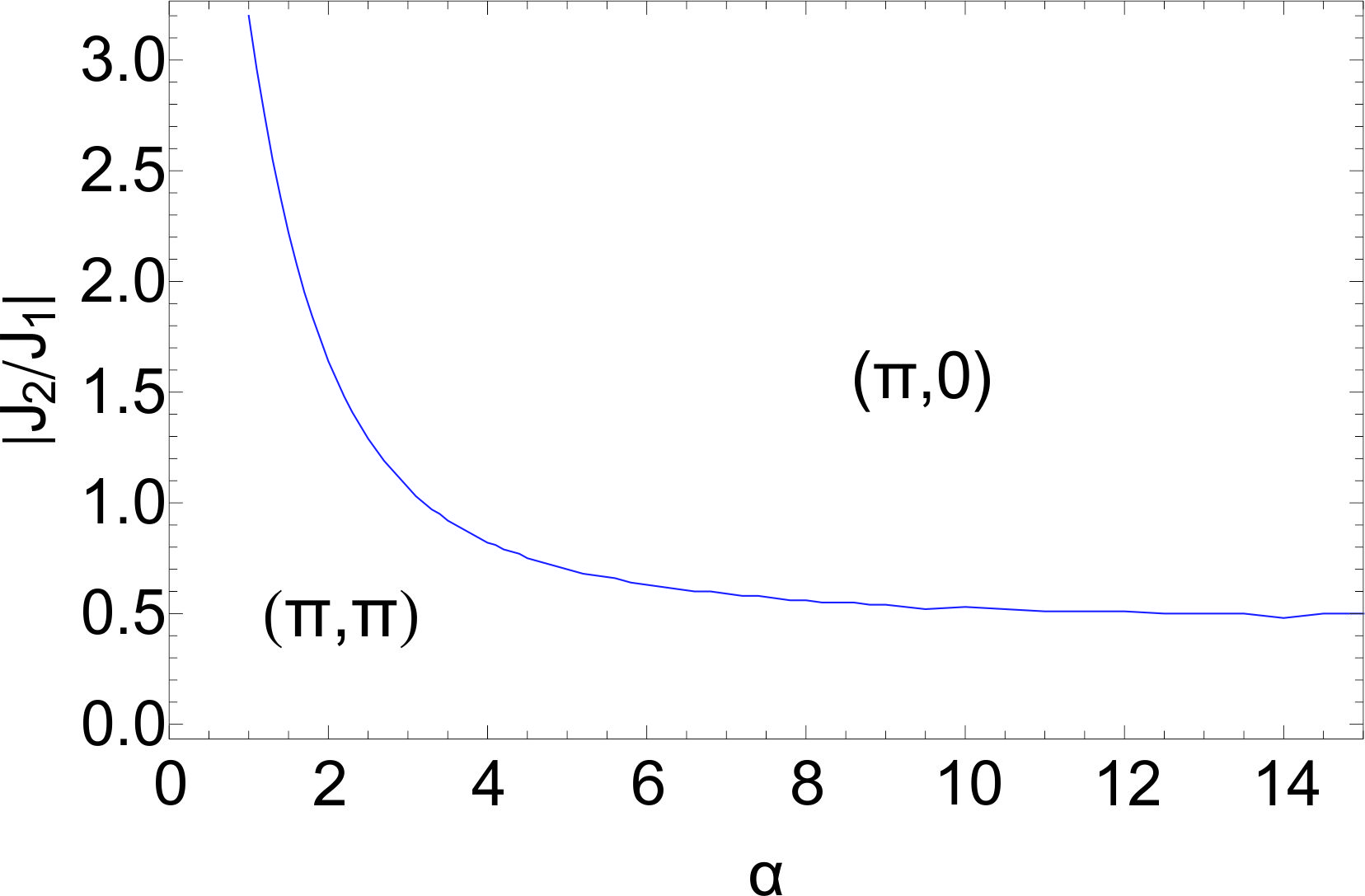}}
\end{center}
\caption{Phase transition point vs. $\alpha$ with $J_1=1$ and $\lambda=1$.} \label{phasechange}
\end{figure}
%%%%%%%%%%%%%%%%%%%%%%%%%%%%%%%%%%%%%%%%%%%%%%%%%%%%%%
In the spin wave theory, both the quantum zero point fluctuations
and thermal fluctuations can reduce the magnetic moment. The sublattice magnetization $m$ is defined as
\begin{equation}
m=<S_i^Z>=S-\Delta m,
\label{eqnm}
\end{equation}
 where $\Delta m$ is the deviation of sublattice magnetization from the saturation value,
\begin{eqnarray}
  \Delta m &=& <a_i^+a_i> \nonumber  \\
           &=& \sum_{\mathbf{k}}
           <a_{\mathbf{k}}^+a_{\mathbf{k}}>\nonumber \\
            &=& \frac{1}{2V_{\mathbf{k}}}\sum_{\mathbf{k}}
           [\frac{SA_{\mathbf{k}}}{\omega(\mathbf{k})}-1]+\frac{1}{V_{\mathbf{k}}}
           \sum_{\mathbf{k}} \frac{SA_{\mathbf{k}}}{\omega(\mathbf{k})}
           \frac{1}{e^{\beta \omega(\mathbf{k})}-1}\nonumber \\
           &=& \Delta m^{quantum} + \Delta m^{thermal}.
\end{eqnarray}
The first term $\Delta m^{quantum}$ comes from the quantum zero
point fluctuations and the second term $\Delta m^{thermal}$
corresponds to the thermal fluctuations. Experimentally, the thermal
fluctuations at low temperature are generally weak and can be
ignored.~\cite{keimer08,boothroyd08,canfield10} In the present
paper, we focus on the quantum zero point fluctuations.

The sublattice magnetization $\Delta m^{quantum}$ can be calculated
by ~\cite{yaofront09}
\begin{equation}
  \Delta m^{quantum} =\frac{1}{2} \int_0^{2\pi}\int_0^{2\pi} \frac{dk_x}{2\pi} \frac{dk_y}{2\pi} \frac{SA_{\mathbf{k}}}{\omega(\mathbf{k})} -\frac{1}{2}.
\end{equation}
It is difficult to get the analytical form of $\Delta m^{quantum}$.
Thus we numerically calculate $\Delta m^{quantum}$ and $m$.

%In Fig.~\ref{m} , $\Delta m^{quantum}$ is plotted as a function of the superexchange coupling ration $|J_2/J_1|$. when $\alpha$ is big enough, for example, $\alpha=10$, which means the long- range interactions is low enough, $\Delta m^{quantum} \rightarrow \infty$ at the point $|J_2/J_1|\approx 0.5$ and the phase transition occures (see Fig~\ref{transitionpipi}). It is known that $\Delta m^{quantum} \rightarrow \infty$ at the point $|J_2/J_1|=0.5$ without long-range interactions. We also find that when $\alpha =4$,the transition point is $|J_2/J_1|=0.825$.It means that the long-range interactions shifts the phase transition point to the right. In addition, $\Delta m^{quantum}$ increases as $|J_2/J_1|$ increases.

%The same things happen in $(\pi,0)$ antiferromagnet (see Fig~\ref{transitionpi0}).$\Delta m^{quantum} \rightarrow \infty$ at the point $|J_2/J_1|=0.52$ with $\alpha =10$  and $\Delta m^{quantum} \rightarrow \infty$ at the point $|J_2/J_1|=0.825$ with $\alpha =4$. We can find that when $\alpha$ is the same, phase transition for $(\pi,\pi)$ antiferromagnet and $(\pi,0)$ antiferromagnet happen at the same point.It means that if $|J_2|$ is big enough, $(\pi,\pi)$ phase translate in to $(\pi,0)$ phase for it has lower energy. We will come back to this point in section phase tansition.On the contrary,$\Delta m^{quantum}$ decreases with the increase of $|J_2/J_1|$.

In Fig.~\ref{alltransition}, $m$ is plotted as a function of
interaction ratio $|J_2/J_1|$ for $S=1$. When $\alpha$ is big
enough, for example, $\alpha=10$, which corresponds to the short
range interactions, $m$ drops to $0$ at the point $|J_2/J_1|\approx
0.5$ and the phase transition happens. This result is consistent
with the $J_1$-$J_2$ model. When long-range interactions are
introduced, the phase transition point shifts to the right. For
example, it becomes $\sim 0.825$ when $\alpha =4$, shown in
Fig.~\ref{alltransition}. From the sharpness of $m$ near the
transition point, we can see that staggered long-range interactions suppresses
quantum fluctuations of spins and enlarges the ordered moment, especially in the Neel state.

\subsection{Phase transition}

%According to experiments, $J_2$ is almost $\sim 0.1$ in the $(\pi,\pi)$ stripe phase, and $\sim 0.8$ in the $(\pi, 0)$ stripe phase. So we use $J_2=0.1$ and $1$ as representatives for discussion.

There is a competition between the $(\pi,\pi)$ Neel phase and $(\pi, 0)$ stripe phase. From the reduced magnetic moment, we get a phase diagram (see Fig.~\ref{phasechange}) for the competing Neel and $(\pi,0)$ stripe states of system containing first- and second-neighbor interactions, in the presence of staggered power-law interactions. The phase transition point is plotted as a function of $\alpha$, which controls the long-range interactions. It can be found that the $(\pi,\pi)$ Neel phase is below the transition line, while the $(\pi, 0)$ stripe phase is above it.
 %If we draw a vertical line across the phase boundary, i.e., $\alpha$ is constant, the $(\pi,\pi)$ Neel phase translates into the $(\pi,0)$ stripe phase with $|J_2/J_1|$ increasing.
As $\alpha$ increases, the phase transition point ($|J_2/J_1|$) decreases quickly and saturates at $0.5$, recovering the result of $J_1$-$J_2$ model. When $\alpha$ approaches $1$, i.e. the interactions decay slowly, the phase transition point $|J_2/J_1|$ increases to $\sim 3.21$ if the longest $J_{(m, n)}=J_{(3, 3)}$. \\

\section{Conclusions}
In conclusion, we have studied the magnetic excitations and phase
transition on the square lattice with long-range interactions, which
are related to the cuprate superconductors and iron-based
superconductors. The general solutions of spin waves have been
worked out for the system with any-order long-range interactions for
the $(\pi, \pi)$ Neel phase and $(\pi, 0)$ stripe phase.
Particulary, for the system with power-law long-range interactions,
we have calculated the spin wave dispersions, spin wave velocities,
dynamic structure factor, constant energy cutting plots, reduced
magnetic moment and phase diagram. The spin wave cones at the $(\pi,
\pi)$ Neel phase are found to be more circular at low energies
because of the existence of long-range interactions. At the $(\pi,
0)$ stripe phase, the spin wave cones are general elliptical at low
energies and the long-range interactions suppress the whole energy
band. At high energies, the long-range interactions have the obvious
effect to the magnetic excitations which can be measured by the
inelastic neutron scattering, NMR, $\mu$SR, etc. The remarked calculated
magnetic moment can be used to examine the effects of long-range
interactions in real materials. We have found surprisingly that staggered long-range interaction can shift the phase transition point and suppress the
quantum fluctuation of spin and enlarges the ordered moment,especially in the Neel state. Our study provides very general results for the two-dimensional
square lattice with long-range interactions, which can be used by
both theoretical and experimental studies.

\begin{acknowledgments}
We thank Wei Ku, Nvsen Ma and Bo Li for helpful discussions. This work is supported by the MOST of China 973 program (2012CB821400), NSFC-11074310, NSFC- Specialized Research Fund for the Doctoral Program of Higher Education (20110171110026), Fundamental Research Funds for the Central Universities of China, and
NCET-11-0547.

\end{acknowledgments}

%\bibliography{bigbib}
\bibliographystyle{forprb}

\end{document}